\documentclass[12pt]{article}
\usepackage{amssym}
\usepackage{graphics}

\def\sech{\mathop{\rm sech}\nolimits}

\title{
New approach to (quasi)-exactly solvable Schr\"odinger equations with a 
position-dependent effective mass}

\author{B.\ Bagchi $^a$, P.\ Gorain $^a$, C.\ Quesne $^{b,}$\thanks{Corresponding
author. {\sl E-mail addresses}: bbagchi123@rediffmail.com (B.\ Bagchi),
psgorain@rediffmail.com (P.\ Gorain), cquesne@ulb.ac.be (C.\ Quesne), raj@isical.ac.in (R.\
Roychoudhury).}\ , R.\ Roychoudhury
$^c$\\ {\small\sl $^a$ Department of Applied Mathematics, University of Calcutta,}\\
{\small\sl 92 Acharya Prafulla Chandra Road, Kolkata 700009, India}\\ 
{\small\sl $^b$ Physique Nucl\'eaire Th\'eorique et Physique Math\'ematique,  Universit\'e
Libre de Bruxelles,} \\  {\small \sl Campus de la Plaine CP229, Boulevard~du Triomphe,
B-1050 Brussels, Belgium}\\
{\small\sl $^c$ Physics and Applied Mathematics Unit, Indian Statistical Institute,
Kolkata 700035, India}\\}
\date{ }
\begin{document}
\baselineskip=22pt plus 1pt minus 1pt
%%%%%%%%%%%%%%%%%%%%%%%%%%%%%%%%%%%%%%%%%%%%%%%%%%%%%%%%%%
\maketitle

\begin{abstract}
By using the point canonical transformation approach in a manner distinct from previous
ones, we generate some new exactly solvable or quasi-exactly solvable potentials for the
one-dimensional Schr\"odinger equation with a position-dependent effective mass. In the
latter case, SUSYQM techniques provide us with some additional new potentials.
\end{abstract}

\vspace{0.5cm}


 
%\bigskip\noindent
%Corresponding author: C.\ Quesne, Physique Nucl\'eaire Th\'eorique et Physique
%Math\'e\-ma\-ti\-que,  Universit\'e Libre de Bruxelles, Campus de la Plaine
%CP229, Boulevard du Triomphe, B-1050 Brussels, Belgium

%\noindent
%Telephone: 32-2-6505559

%\noindent
%Fax: 32-2-6505045

%\noindent
%E-mail: cquesne@ulb.ac.be 
\newpage
%
%========================================================================
%
\section{Introduction}

In recent years, quantum mechanical systems with a position-dependent effective mass
(PDEM) have attracted a lot of attention due to their relevance in describing the physics
of many microstructures of current interest, such as compositionally graded
crystals~\cite{geller}, quantum dots~\cite{serra}, ${}^3$He clusters~\cite{barranco},
quantum liquids~\cite{arias}, metal clusters~\cite{puente}, etc.\par
%
%----------------------------------------------------------------------------------------
%
As in the constant-mass case, exact solutions play an important role because they may
provide both a conceptual understanding of some physical phenomena and a testing
ground for some approximation schemes. Many recent developments have been devoted
to constructing exactly solvable (ES), quasi-exactly solvable (QES) or conditionally-exactly
solvable potentials for the PDEM Schr\"odinger equation \cite{dekar}--\cite{bagchi05} by
using point canonical transformations (PCT), Lie algebraic techniques or supersymmetric
quantum mechanical (SUSYQM) methods.\par
%
%-------------------------------------------------------------------------------------
%
In this Letter, we will show that new ES or QES potentials in a PDEM background
may be generated by using the PCT approach in a manner distinct from previous ones.
We will then combine such results  with SUSYQM methods to produce some additional
QES potentials.\par
%
%===============================================
% 
\section{PCT approach in a PDEM context}

As is well known (see, e.g., \cite{bagchi04a}), the general Hermitian PDEM Hamiltonian,
initially proposed by von Roos~\cite{vonroos} in terms of three ambiguity parameters
$\alpha$, $\beta$, $\gamma$ such that $\alpha + \beta + \gamma = -1$, gives rise to
the (time-independent) Schr\"odinger equation
\begin{equation}
  H \psi(x) \equiv \left[- \frac{d}{dx} \frac{1}{M(x)} \frac{d}{dx} + V_{\rm eff}(x)
  \right] \psi(x) = E \psi(x),  \label{eq:H}
\end{equation}
where the effective potential
\begin{equation}
  V_{\rm eff}(x) = V(x) + \frac{1}{2} (\beta + 1) \frac{M''}{M^2} - [\alpha (\alpha +
  \beta + 1) + \beta + 1] \frac{M^{\prime 2}}{M^3}  \label{eq:Veff} 
\end{equation}
depends on some mass terms. Here a prime denotes derivative with respect to $x$,
$M(x)$ is the dimensionless form of the mass function $m(x) = m_0 M(x)$ and we have
set $\hbar = 2m_0 = 1$.\par
%
%----------------------------------------------------------------------------------
%
Let us look for solutions of Eq.~(\ref{eq:H}) of the form
\begin{equation}
  \psi(x) = f(x) F(g(x)),  \label{eq:psi}
\end{equation}
where $f(x)$, $g(x)$ are two so far undetermined functions and $F(g)$ satisfies a
second-order differential equation
\begin{equation}
  \ddot{F} + Q(g) \dot{F} + R(g) F = 0,  \label{eq:eq-F}
\end{equation}
where a dot denotes derivative with respect to $g$. Since in this Letter we shall be
interested in bound-state wavefunctions, we shall actually restrict ourselves to polynomial
solutions of Eq.~(\ref{eq:eq-F}).\par
%
%--------------------------------------------------------------------------
%
On inserting Eq.~(\ref{eq:psi}) in Eq.~(\ref{eq:H}) and comparing the result with
Eq.~(\ref{eq:eq-F}), we arrive at two expressions for $Q(g(x))$ and $R(g(x))$ in terms of
$E - V_{\rm eff}(x)$ and of $M(x)$, $f(x)$, $g(x)$ and their derivatives. The former
allows us to calculate $f(x)$, which is given by
\begin{equation}
  f(x) \propto \left(\frac{M}{g'}\right)^{1/2} \exp\left(\frac{1}{2} \int^{g(x)} Q(u)\, 
  du\right),  \label{eq:f}
\end{equation}
while the latter leads to the equation
\begin{equation}
  E - V_{\rm eff}(x) = \frac{g'''}{2Mg'} - \frac{3}{4M} \left(\frac{g''}{g'}\right)^2 +
  \frac{g^{\prime2}}{M} \left(R - \frac{1}{2} \dot{Q} - \frac{1}{4} Q^2\right) - 
  \frac{M''}{2M^2} + \frac{3 M^{\prime2}}{4 M^3}.  \label{eq:PCT}
\end{equation}
It is clear that we need to find some functions $M(x)$, $g(x)$ ensuring the presence of a
constant term on the right-hand side of Eq.~(\ref{eq:PCT}) to compensate $E$ on its
left-hand side and giving rise to an effective potential $V_{\rm eff}(x)$ with well-behaved
wavefunctions.\par
%
%---------------------------------------------------------------------------
%
In the constant-mass case, i.e., for $M(x) = 1$, this procedure has been thoroughly
investigated~\cite{bhatta, levai}. A similar study in the PDEM context looks more involved
for two reasons: (i) there are now two unknown functions instead of only one and (ii) the
usual square-integrability condition for bound-state wavefunctions has to be completed
by the additional restriction $|\psi(x)|^2/\sqrt{M(x)} \to 0$ at the end points of the
definition interval of $V(x)$ to ensure the Hermiticity of $H$ in the Hilbert space
spanned by its eigenfunctions~\cite{bagchi05}.\par
%
%---------------------------------------------------------------------------------
% 
In most applications of PCT that have been carried out so far in the PDEM context, the
choice $M = \lambda g^{\prime2}$ or $g(x) = (1/\lambda) \int^x \sqrt{M(u)}\, du +
\nu$ (where $\lambda$, $\nu$ are some constants) has been made (see, e.g.,
\cite{alhaidari, gonul, bagchi04a}). In the next two sections, we will explore the new
possibilities offered by two other options, namely $M = \lambda g'$ and $M =
\lambda/g'$ or, equivalently, $g(x) = (1/\lambda) \int^x M(u)\, du + \nu$ and $g(x) = 
(1/\lambda) \int^x [M(u)]^{-1}\, du + \nu$.\par
%
%============================================
%
\section{\boldmath Generation of ES potentials in the $M = \lambda g'$ case}

Substituting $M = \lambda g'$ into Eq.~(\ref{eq:PCT}) leads to
\begin{equation}
  E - V_{\rm eff}(x) = \frac{1}{\lambda} g' \left(R - \frac{1}{2} \dot{Q} - \frac{1}{4}
  Q^2\right).  \label{eq:PCT-ES}
\end{equation}
Some simple and interesting results can be derived from this relation by assuming that
$F(g)$ is either a Jacobi or a generalized Laguerre polynomial~\cite{abramowitz}.\par
%
%------------------------------------------------------------------------------
% 
{}For $F_n(g) \propto P^{(a,b)}_n(g)$, $n=0$, 1, 2,~\ldots, $a, b > -1$, we
obtain
\begin{eqnarray}
  R - \frac{1}{2} \dot{Q} - \frac{1}{4} Q^2 & = & \frac{n(n+a+b+1)}{1-g^2} + \frac{1}
        {(1-g^2)^2} \left[\frac{1}{2} (a+b+2) - \frac{1}{4} (b-a)^2\right] \nonumber \\
  && \mbox{} + \frac{g}{(1-g^2)^2} \frac{1}{2} (b-a)(b+a) - \frac{g^2}{(1-g^2)^2}
        \frac{1}{4} (a+b)(a+b+2).  \label{eq:R-Jacobi}  
\end{eqnarray}
A constant term can therefore be generated on the right-hand side of
Eq.~(\ref{eq:PCT-ES}) by assuming $g'/[\lambda (1-g^2)] = C$, where $C$ must be
restricted to positive values in order to get increasing energy eigenvalues for successive
$n$ values. The solution of this first-order differential equation for $g(x)$ leading to a
positive mass function reads $g(x) = \tanh qx$, where $q = \lambda C > 0$. Without loss
of generality, we may set $C = q^2$ so that $\lambda = 1/q$. Hence we get
\begin{equation}
  g(x) = \tanh qx, \qquad M(x) = \sech^2 qx, \qquad - \infty < x < + \infty.
  \label{eq:g-Jacobi}
\end{equation}
\par
%
%-----------------------------------------------------------------------------------
%
Equations (\ref{eq:f}), (\ref{eq:PCT-ES}), (\ref{eq:R-Jacobi}) and (\ref{eq:g-Jacobi}) then
yield
\begin{eqnarray}
  E_n & = & q^2 \left(n + \frac{a+b}{2}\right) \left(n + \frac{a+b+2}{2}\right) + V_0, 
      \label{eq:E-Jacobi} \\
  V_{\rm eff}(x) & = & q^2 \left\{\left[\frac{1}{2} (a^2+b^2) - 1\right] \cosh^2 qx
      + \frac{1}{2} (a-b)(a+b) \sinh qx \cosh qx\right\} + V_0 \nonumber \\
  & = & \frac{1}{4} q^2 \left[(a^2-1) e^{2qx} + (b^2-1) e^{-2qx} + a^2 + b^2 - 2\right]
      + V_0, \\
  \psi_n(x) & \propto & (1 - \tanh qx)^{(a+1)/2} (1 + \tanh qx)^{(b+1)/2} P^{(a,b)}_n
      (\tanh qx), \label{eq:psi-Jacobi}
\end{eqnarray}
where $n=0$, 1, 2,~\ldots, $V_0$ denotes some constant and we have to assume $a, b
> -1/2$ in order to satisfy the conditions on bound-state wavefunctions in a PDEM
context (observe that the square-integrability condition alone does not impose any
restriction on $a$, $b$!).\par
%
%---------------------------------------------------------------------------------
%
By proceeding similarly for $F_n(g) \propto L^{(a)}_n(g)$, $n=0$, 1, 2,~\ldots, $a >
-1$, from the relation~\cite{abramowitz}
\begin{equation}
  R - \frac{1}{2} \dot{Q} - \frac{1}{4} Q^2 = \frac{2n+a+1}{2g} - \frac{(a+1)(a-1)}
  {4g^2} - \frac{1}{4}
\end{equation}
and the condition $g'/(\lambda g) = C > 0$, we obtain the results
\begin{equation}
  g(x) = e^{-qx}, \qquad M(x) = e^{-qx}, \qquad - \infty < x < + \infty,
\end{equation}
where we have set $C = q^2$ (hence $\lambda  = - 1/q$) and where without loss of
generality we may assume $q > 0$. Furthermore
\begin{eqnarray}
  E_n & = & q^2 \left(n + \frac{a+1}{2}\right) + V_0, \\
  V_{\rm eff}(x) & = & \frac{1}{4} q^2 \left[(a^2-1) e^{qx} + e^{-qx} \right] + V_0, \\
  \psi_n(x) & \propto & \exp\left\{- \frac{1}{2} \left[(a+1)qx + e^{-qx}\right]\right\}
       L^{(a)}_n\left(e^{-qx}\right),
\end{eqnarray}
where the PDEM background imposes an additional restriction $a > - 1/2$ on the
wavefunctions again.\par
%
%--------------------------------------------------------------------------------
%
Turning now to the initial potential $V(x)$, we find from Eq.~(\ref{eq:Veff}) that $V(x) =
V_{\rm eff}(x) + q^2 [f(\alpha, \beta) \cosh^2 qx - g(\alpha, \beta)]$ and $V(x) =
V_{\rm eff}(x) + \frac{1}{4} q^2 f(\alpha, \beta) e^{qx}$, with $f(\alpha, \beta) \equiv
(2\alpha+1) (2\alpha+2\beta+2) - 2\alpha$, $g(\alpha, \beta) \equiv (2\alpha+1)^2 +
\beta (4\alpha+1)$, for the Jacobi  and generalized Laguerre polynomials, respectively.
Hence, in both cases, for the choice of ambiguity parameters made by BenDaniel and Duke
($\alpha=0$, $\beta=-1$)~\cite{bendaniel}, there is no distinction between $V(x)$ and
$V_{\rm eff}(x)$. Furthermore, when the Jacobi polynomials reduce to Legendre ones,
i.e., for $a=b=0$, and the ambiguity parameters are those selected by Zhu and Kroemer
($\alpha = - 1/2$, $\beta=0$)~\cite{zhu}, $V(x)$ becomes a constant potential
$V_0$. Our results (\ref{eq:E-Jacobi}) and (\ref{eq:psi-Jacobi}) then describe the
generation of an infinite number of bound states for a free-particle potential in a
$\sech^2$-mass environment~\cite{bagchi04a}. For nonvanishing $a$, $b$ values,
Eqs.~(\ref{eq:E-Jacobi})--(\ref{eq:psi-Jacobi}) may therefore be seen as a generalization
of this interesting property.\par
%
%=============================================
% 
\section{\boldmath Generation of QES potentials in the $M = \lambda/g'$ case}

Whenever $M = \lambda/g'$, Eq.~(\ref{eq:PCT}) becomes
\begin{equation}
  E - V_{\rm eff}(x) = \frac{g'''}{\lambda} - \frac{g^{\prime\prime2}}{\lambda g'} +
  \frac{g^{\prime3}}{\lambda} \left(R - \frac{1}{2} \dot{Q} - \frac{1}{4} Q^2\right).
  \label{eq:PCT-QES}
\end{equation}
In such a case, we shall take for $F(g)$ some polynomials of nonhypergeometric type
satisfying the equation 
\begin{equation}
  \ddot{F} + \frac{a (g^2 - \xi^2)}{g^3} \dot{F} + \frac{bg+c}{g^3} F = 0,
  \label{eq:krylov}
\end{equation}
where we assume $a$, $b$, $c$, $\xi$ real, $a \ne 0$, $b \ne 0$ and $\xi > 0$. As shown
elsewhere~\cite{krylov}, this second-order differential equation has $k$th-degree
polynomial solutions provided $b = -k (a+k-1)$ and there exist $k+1$ such solutions
$F_n(g)$, $n=0$, 1,~\ldots, $k$, associated with $k+1$ distinct values $c_n$ of $c$, if
$a$ is appropriately chosen.\par
%
%-----------------------------------------------------------------------------
%
Substituting
\begin{equation}
  R - \frac{1}{2} \dot{Q} - \frac{1}{4} Q^2 = - \frac{(2k+a-2)(2k+a)}{4g^2} + \frac{c_n}
  {g^3} + \frac{a(a-3)\xi^2}{2g^4} - \frac{a^2\xi^4}{4g^6}
\end{equation}
in Eq.~(\ref{eq:PCT-QES}), we find a constant term on the right-hand side of the
transformed equation by choosing $g^{\prime3}/(\lambda g^3) = C$. Then with
$C=q^2$ and $\lambda = q > 0$, we obtain
\begin{equation}
  g(x) = e^{qx}, \qquad M(x) = e^{-qx}, \qquad - \infty < x < + \infty. \label{eq:g-QES}
\end{equation}
Hence
\begin{eqnarray}
  E_n & = & q^2 c_n + V_0, \\
  V_{\rm eff}(x) & = & q^2 \left[\frac{1}{4} (2k+a-2)(2k+a) e^{qx} - \frac{1}{2} a(a-3)
       \xi^2 e^{-qx} + \frac{1}{4} a^2 \xi^4 e^{-3qx}\right] + V_0, \label{eq:Veff-QES} \\
  \psi_n(x) & \propto & \exp\left[\frac{1}{2}(a-2)qx + \frac{1}{4} a \xi^2 e^{-2qx}
       \right] F_n\left(e^{qx}\right), \label{eq:psi-QES}
\end{eqnarray}
where $n=0$, 1,~\ldots, $k$.\par
%
%-------------------------------------------------------------------------------
%
The functions (\ref{eq:psi-QES}) turn out to be physically acceptable as bound-state
wavefunctions provided $a$ is restricted to the range $a < - 2k + \frac{3}{2}$. We
conclude that for such values and for the PDEM given in (\ref{eq:g-QES}), the effective
potentials (\ref{eq:Veff-QES}) corresponding to $k=1$, 2, 3,~\ldots, are QES with $k+1$
known eigenvalues and eigenfunctions. For $k=1$ and $k=2$, for instance, we find
\begin{equation}
  c_{\stackrel{0}{1}} = \pm a \xi, \qquad F_{\stackrel{0}{1}}(g) \propto g \pm \xi,
  \qquad {\rm if\ } a < - \frac{1}{2},
\end{equation}
and 
\begin{eqnarray}
  c_{\stackrel{0}{2}} & = & \mp \Delta \xi, \qquad c_1 = 0, \qquad
       F_{\stackrel{0}{2}}(g) \propto g^2 \mp \frac{\Delta}{a+2} \xi g + \frac{a}{a+2}
       \xi^2, \nonumber \\
  F_1(g) & \propto & g^2 - \frac{a}{a+1} \xi^2, \qquad \Delta \equiv \sqrt{2a(2a+3)},
       \qquad {\rm if\ } a < - \frac{5}{2}, 
\end{eqnarray}
respectively. Observe on these two examples that for the values taken by $a$, $\xi$ and
$g(x)$, $\psi_n(x)$ has $n$ zeros on the real line, so that $\psi_0(x)$ is the
ground-state wavefunction, while $\psi_n(x)$, $n=1$, 2,~\ldots, $k$, correspond to the
$n$th excited states.\par
%
%----------------------------------------------------------------------------
% 
The results presented here could be easily extended to more general polynomials. For
instance, if instead of $g^2 - \xi^2$ in (\ref{eq:krylov}), we had considered $(g - g_3)(g -
g_4)$ with $g_3$ and $g_4$ real but $g_4 \ne - g_3$, we would have obtained 
effective potentials containing an additional term proportional to $e^{-2qx}$.\par
%
%---------------------------------------------------------------------
%
{}Finally, it should be noticed that the PDEM being the same as that chosen for
generalized Laguerre polynomials in Sec.~3, the relation between $V(x)$ and $V_{\rm
eff}(x)$ is also similar.\par
%
%====================================================
%
\section{SUSYQM approach}

Let us consider the intertwining relationship $\eta H = H_1 \eta$, where $H$ is the
Hamiltonian defined in Eq.~(\ref{eq:H}), $H_1$ has the same kinetic energy term but an
associated effective potential $V_{1,{\rm eff}}(x)$, and $\eta$ is a first-order
intertwining operator $\eta = A(x) \frac{d}{dx} + B(x)$. As shown in \cite{bagchi04a},
such a relationship leads to the restrictions $A(x) = M^{-1/2}$ and
\begin{equation}
  V_{\rm eff}(x) = \epsilon + B^2 - \left(\frac{B}{\sqrt{M}}\right)', \qquad 
  V_{1,{\rm eff}}(x) = V_{\rm eff} + \frac{2B'}{\sqrt{M}} + \frac{M''}{2M^2} -
  \frac{3M^{\prime2}}{4M^3},  \label{eq:partner-V}
\end{equation}
with $\epsilon$ denoting some arbitrary constant.\par
%
%------------------------------------------------------------------------------
%
A solution for $B(x)$, which at the same time ensures that $\eta$ annihilates the
ground-state wavefunction of $H$, is provided by $B(x) = - \psi'_0/(\sqrt{M} \psi_0)$
together with $\epsilon = E_0$. In this (PDEM-extended) unbroken SUSYQM
framework~\cite{cooper}, the eigenvalues of $H_1$ are $E_{1,n} = E_{n+1}$, $n=0$, 1,
2,~\ldots, with the corresponding wavefunctions given by $\psi_{1,n} \propto \eta
\psi_{n+1}$.\par
%
%-----------------------------------------------------------------------------------------
%
{}For the wavefunctions considered in Eq.~(\ref{eq:psi}), $\psi'_0/\psi_0$ in general
contains two terms: $\psi'_0/\psi_0 = f'/f + g' \dot{F}_0/F_0$. In the ES potential case
reviewed in Sec.~3, however, the second term vanishes since $F_0(g) = 1$, so that we
obtain simple results for $B(x)$, namely
\begin{equation}
  B(x) = \frac{1}{2}q [(a-b) \cosh qx + (a+b+2) \sinh qx]
\end{equation}
and
\begin{equation}
  B(x) = \frac{1}{2}q [(a+1) e^{qx/2} - e^{-qx/2}]
\end{equation}
for the Jacobi and generalized Laguerre polynomials, respectively. Substituting such
functions in (\ref{eq:partner-V}), we arrive at SUSY partners $V_{1,{\rm eff}}(x)$, which
have the same shape as $V_{\rm eff}(x)$ and differ only in the parameters ($a_1 =
a+1$, $b_1 = b+1$, $V_{0,1} = V_0$ and $a_1 = a+1$, $V_{0,1} = V_0 + \frac{1}{2}
q^2$, respectively). We conclude that the potentials $V_{\rm eff}(x)$ are shape
invariant.\par
%
%------------------------------------------------------------------------------
%
The QES potential case reviewed in Sec.~4 looks more interesting because $F_0(g)$
being now a $k$th-degree polynomial in $g$, the second term in $\psi'_0/\psi_0$ does
not vanish anymore. As a consequence, the functions $B(x)$ and $V_{1,{\rm eff}}(x)$
become $k$-dependent and given by
\begin{eqnarray}
  B(x) & = & q \left[- \frac{1}{2}(a-2) e^{qx/2} + \frac{1}{2}a \xi^2 e^{-3qx/2} - 
       e^{3qx/2} \frac{\dot{F}_0}{F_0}\right], \\
  V_{1,{\rm eff}}(x) & = & V_{\rm eff} - q^2 \left[\frac{1}{2} \left(a - \frac{3}{2}
       \right) e^{qx} + \frac{3}{2}a\xi^2 e^{-qx} + 3 e^{2qx} \frac{\dot{F}_0}{F_0} + 2
       e^{3qx} \left(\frac{\ddot{F}_0}{F_0} - \frac{\dot{F}_0^2}{F_0^2}\right)\right]. 
\end{eqnarray}
The SUSY partners $V_{1,{\rm eff}}(x)$ therefore contain some terms which
are rational functions in $e^{qx}$. For $k=1$ and $k=2$, for instance, we obtain
\begin{eqnarray}
  V_{1,{\rm eff}}(x) & = & q^2 \Biggl[\frac{1}{4}(a-1)(a+1) e^{qx} - \frac{1}{2}a^2
       \xi^2 e^{-qx} + \frac{1}{4} a^2 \xi^4 e^{-3qx} \nonumber \\
  && \mbox{} + \frac{3\xi^2}{e^{qx}+\xi} - \frac{2\xi^3}{(e^{qx}+\xi)^2} \Biggl] +
       V_0 - q^2 \xi 
\end{eqnarray}
and
\begin{eqnarray}
  V_{1,{\rm eff}}(x) & = & q^2 \Biggl[\frac{1}{4}(a+1)(a+3) e^{qx} - \frac{1}{2}a^2
       \xi^2 e^{-qx} + \frac{1}{4} a^2 \xi^4 e^{-3qx} \nonumber \\
  && \mbox{} + \frac{a\xi^2}{(a+2)^3} Z_1(x) - \frac{4a^2\xi^4}{(a+2)^4} Z_2(x)
       \Biggl] + V_0 + q^2 \frac{\Delta\xi}{a+2}, \nonumber \\
  Z_1(x) & \equiv & \frac{6(a+2)(a+1)e^{qx} - (a+6)\Delta\xi}{e^{2qx} - \frac{\Delta}
       {a+2}\xi e^{qx} + \frac{a}{a+2}\xi^2}, \nonumber \\
  Z_2(x) & \equiv & \frac{(3a+4)e^{qx} - \Delta\xi}{\left(e^{2qx} - \frac{\Delta}
       {a+2}\xi e^{qx} + \frac{a}{a+2}\xi^2\right)^2}, 
\end{eqnarray}
respectively. Such effective potentials provide us with some new examples of QES
potentials in a PDEM environment with $k$ known eigenvalues and eigenfunctions.\par
%
%============================================
%
\section{Conclusion}

In this Letter, we have investigated the problem of the one-dimensional Schr\"odinger
equation in a PDEM background from several viewpoints. By using first the PCT approach
and assuming a relation between the new variable $g = g(x)$ and the mass $M(x)$ that
differs from the usual one, we have constructed some new ES or QES potentials. The
former are associated with either Jacobi or generalized Laguerre polynomials, while the
latter correspond to some $k$th-degree polynomials of nonhypergeometric type.\par
%
%----------------------------------------------------------------------------
%
We have then considered an equivalent intertwining-operator approach and shown that
while our ES potentials are shape invariant, the SUSY partners of our QES potentials are
new. In the latter case, iterating the procedure would lead us to a hierarchy of SUSY
partners with an increasingly complicated form.\par
%
%---------------------------------------------------------------------------
%
The method described here could be used to generate other classes of masses and
potentials providing exact solutions of the PDEM Schr\"odinger equation.\par
%
%===========================================
%
\section*{Acknowledgements}

PG thanks the Council of Scientific and Industrial Research (CSIR), New Delhi for the award
of a fellowship. CQ is a Research Director of the National Fund for Scientific Research
(FNRS), Belgium.\par
%
%=========================================================
%


\newpage
\begin{thebibliography}{99}

\bibitem{geller} M.R.\ Geller, W.\ Kohn, Phy.\ Rev.\ Lett.\ 70 (1993) 3103.

\bibitem{serra} Ll.\ Serra, E.\ Lipparini, Europhys.\ Lett.\ 40 (1997) 667.

\bibitem{barranco} M.\ Barranco, M.\ Pi, S.M.\ Gatica, E.S.\ Hern\'andez, J.\ Navarro, 
Phys.\ Rev.\ B 56 (1997) 8997.

\bibitem{arias} F.\ Arias de Saavedra, J.\ Boronat, A.\ Polls, A.\ Fabrocini, Phys.\ Rev.\
B 50 (1994) 4248.

\bibitem{puente} A.\ Puente, Ll.\ Serra, M.\ Casas, Z.\ Phys.\ D 31 (1994) 283.

\bibitem{dekar} L.\ Dekar, L.\ Chetouani, T.F.\ Hammann, J.\ Math.\ Phys.\ 39 (1998)
2551.

\bibitem{milanovic} V.\ Milanovi\'c, Z.\ Ikoni\'c, J.\ Phys.\ A 32 (1999) 7001.

\bibitem{plastino} A.R.\ Plastino, A.\ Rigo, M.\ Casas, F.\ Garcias, A.\ Plastino, Phys.\
Rev.\ A 60 (1998) 4318.

\bibitem{dutra} A.\ de Souza Dutra, C.A.S.\ Almeida, Phys.\ Lett.\ A 275 (2000) 25.

\bibitem{roy} B.\ Roy, P.\ Roy, J.\ Phys.\ A 35 (2002) 3961.

\bibitem{koc} R.\ Ko\c c, M.\ Koca, E.\ K\"orc\"uk, J.\ Phys.\ A 35 (2002) L527; \\
R.\ Ko\c c, M.\ Koca, J.\ Phys.\ A 36 (2003) 8105.

\bibitem{alhaidari} A.D.\ Alhaidari, Phys.\ Rev.\ A 66 (2002) 042116.

\bibitem{gonul} B.\ G\"on\"ul, O.\ \"Ozer, B.\ G\"on\"ul, F.\ \"Uzg\"un, Mod.\ Phys.\
Lett.\ A 17 (2002) 2453.

\bibitem{bagchi04a} B.\ Bagchi, P.\ Gorain, C.\ Quesne, R.\ Roychoudhury, Mod.\ Phys.\
Lett.\ A 19 (2004) 2765.

\bibitem{bagchi04b} B.\ Bagchi, P.\ Gorain, C.\ Quesne, R.\ Roychoudhury, Czech.\
J.\ Phys.\ 54 (2004) 1019.

\bibitem{cq} C.\ Quesne, V.M.\ Tkachuk, J.\ Phys.\ A 37 (2004) 4267.

\bibitem{bagchi05} B.\ Bagchi, A.\ Banerjee, C.\ Quesne, V.M.\ Tkachuk, J.\ Phys.\ A 38
(2005) 2929.

\bibitem{vonroos} O.\ von Roos, Phys.\ Rev.\ B 27 (1983) 7547.

\bibitem{bhatta} A.\ Bhattacharjie, E.C.G.\ Sudarshan, Nuovo Cimento 25 (1962) 864; \\
G.A.\ Natanzon, Theor.\ Math.\ Phys.\ 38 (1979) 146.

\bibitem{levai} G.\ L\'evai, J.\ Phys.\ A 22 (1989) 689; \\
G.\ L\'evai, J.\ Phys.\ A 24 (1991) 131; \\
R.\ Roychoudhury, P.\ Roy, M.\ Znojil, G.\ L\'evai, J.\ Math.\ Phys.\ 42 (2001) 1996;\\
B.\ Bagchi and A.\ Ganguly, J.\ Phys.\ A 36 (2003) L161.

\bibitem{abramowitz} M.\ Abramowitz, I.A.\ Stegun, Handbook of Mathematical Functions,
Dover, New York, 1965.

\bibitem{bendaniel} D.J.\ BenDaniel, C.B.\ Duke, Phys.\ Rev.\ B 152 (1966) 683.

\bibitem{zhu} Q.-G.\ Zhu, H.\ Kroemer, Phys.\ Rev.\ B 27 (1983) 3519.

\bibitem{krylov} G.\ Krylov, M.\ Robnik, J.\ Phys.\ A 34 (2001) 5403.

\bibitem{cooper} F.\ Cooper, A.\ Khare, U.\ Sukhatme, Phys.\ Rep.\ 251 (1995) 267.

\end {thebibliography}
\end{document}